\RequirePackage[2020-02-02]{latexrelease}
\documentclass[aps,prl,twocolumn,superscriptaddress]{revtex4}

\usepackage{graphicx}
\usepackage{dcolumn}
\usepackage{amsmath}
\usepackage{enumerate,amsthm,amssymb,color}
\usepackage{bbm}
\usepackage{bm}
\usepackage[english]{babel}
\usepackage{multirow}

\usepackage{natbib}
\setcitestyle{super,sort&compress}
\begin{document}

\preprint{APS/123-QED}

\title{Experimental Demonstration of Discrete Modulation Formats \\ for Continuous Variable Quantum Key Distribution}

\author{Francois Roumestan} \affiliation{Nokia Bell Labs, Paris-Saclay, route de Villejust, F-91620 Nozay, France}\affiliation{Sorbonne Universit\'e, CNRS, LIP6, 4 place Jussieu, F-75005 Paris, France}
\author{Amirhossein Ghazisaeidi} \affiliation{Nokia Bell Labs, Paris-Saclay, route de Villejust, F-91620 Nozay, France}
\author{J\'er\'emie Renaudier} \affiliation{Nokia Bell Labs, Paris-Saclay, route de Villejust, F-91620 Nozay, France}
\author{Luis Trigo Vidarte} \affiliation{ICFO - Institut de Ciències Fotòniques, The Barcelona Institute of Science and Technology, Castelldefels (Barcelona) 08860, Spain}
\author{Anthony Leverrier}\affiliation{Inria  Paris,  2  rue  Simone  Iff,  F75589  Paris  Cedex 12,  France}
\author{Eleni Diamanti}\affiliation{Sorbonne Universit\'e, CNRS, LIP6, 4 place Jussieu, F-75005 Paris, France}
\author{Philippe Grangier}\affiliation{Universit\'e Paris-Saclay, IOGS, CNRS, Laboratoire Charles Fabry, F-91127 Palaiseau, France}

\maketitle

{\bf Quantum key distribution (QKD) enables the establishment of secret keys between users connected via a channel vulnerable to eavesdropping, with information-theoretic security, that is, independently of the power of a malevolent party~\cite{SBC:rmp09}. QKD systems based on the encoding of the key information on continuous variables (CV), such as the values of the quadrature components of coherent states~\cite{WPG:rmp12,DL15}, present the major advantage that they only require standard telecommunication technology. However, the most general security proofs for CV-QKD required until now the use of Gaussian modulation by the transmitter, complicating practical implementations~\cite{JKL+13,ZCP+:prl20,JCM21}. Here, we experimentally implement a protocol that allows for arbitrary, Gaussian-like, discrete modulations, whose security is based on a theoretical proof that applies very generally to such situations~\cite{denys21}.  These modulation formats are compatible with the use of powerful tools of coherent optical telecommunication, allowing our system to reach a performance of tens of megabit per second secret key rates over 25 km.}

\vskip 2mm

Driven by the pressing need for high-security solutions to address risks to cybersecurity posed by rapid technological progress, the development of quantum key distribution (QKD) systems has advanced significantly in recent years~\cite{DLQY:npjQI16,pir2020,xu2020}. A major challenge in this direction is to leverage the high potential of techniques that have been developed with great success for the classical telecommunication industry, with the goal of both enhancing the performance of QKD systems and assuring their smooth integration into deployed fibre optic network infrastructures. Continuous-variable (CV) QKD schemes~\cite{GVW:nat03,DL15} are particularly well suited for this purpose. The key feature of such schemes is that dedicated photon-counting technology required in standard single-photon based schemes can be replaced by coherent detection techniques that are widely used in classical optical communications. This hardware simplification, however, comes at the price of a more involved theoretical analysis, and security proofs typically require the transmitter, commonly called Alice, to prepare coherent states with a Gaussian modulation to be sent to the receiver, Bob. Such a modulation has been used for advanced experimental implementations~\cite{JKL+13,ZCP+:prl20,JCM21}, but is not a common industrial practice;
a more practical approach is to send coherent states chosen from a finite constellation in phase space. Although such discrete modulations were considered early in CV-QKD~\cite{LRH+:pra06,LG09,LG11}, sound security proofs have been developed only recently, for protocols with either very large constellation sizes \cite{KGW21} or very small ones~\cite{GGDL19,LUL19,LL20,MMS21}, with some experimental implementations in the latter case~\cite{HIM+:qst17,WLL+:arXiv21}. But the most interesting format of medium-size quadrature amplitude modulation (QAM) used in classical optical communications remained out of reach for these methods, which rely on solving huge convex optimization problems. This outstanding issue was solved in Ref.~\cite{denys21}, which provided an analytical bound for the asymptotic secret key rate of protocols with arbitrary modulation schemes, including probabilistic constellation shaping (PCS) QAM~\cite{GFR17}. Strictly speaking, this bound is not tight, but it becomes essentially so for any QAM of size greater than 64.

Here, we experimentally demonstrate CV-QKD with PCS 64-QAM and 256-QAM that can reach very high peak secret key rate (SKR) with standard hardware and software compatible with current telecommunication systems~\cite{ECOC2021,PWS:ol22}. We emphasize that our choice of modulation format presents a number of advantages in practice: the use of QAM ensures the need for a smaller number of random numbers and leads in principle to more efficient post-processing, pulse shaping requires a smaller bandwidth, and PCS optimizes the mutual information bringing it closer to Shannon channel capacity. Our results thus open the way towards integrating CV-QKD in standard optical communication systems, in an efficient, transparent, and cost-efficient way.

\paragraph{CV-QKD protocol and security proof.}
In the Prepare-and-Measure (PM) coherent state CV-QKD protocol with discrete modulation, Alice prepares coherent states $|\alpha\rangle =|(p + iq)/2\rangle$, chosen at random from a discrete constellation. She sends them through an optical link to Bob who measures them using coherent detection.
This quantum transmission phase is followed by classical post-processing, in which Alice and Bob compare a randomly chosen fraction of their data to estimate the channel parameters and thus the length of the final key. Then they correct errors through a reconciliation step and finally turn their identical data set into a shorter secret key via privacy amplification.

The security of this PM protocol is analysed through an equivalent Entanglement-Based (EB) protocol~\cite{DL15}, where Alice (virtually) prepares an initial entangled state, measures one mode and transmits the second mode to Bob through the quantum channel. Exploiting the property that Gaussian states maximize the Holevo information between Bob's measurement outcome and the eavesdropper quantum memory~\cite{WGC06,GC:prl06,NGA:prl06}, it is sufficient to compute the covariance matrix of the bipartite state shared by Alice and Bob before measurement. The difficulty is that this virtual state is never prepared nor measured in the true PM protocol. Rather, the goal is for Alice and Bob to infer this covariance matrix from the data they observe in the PM protocol.

While this task is straightforward when the modulation is Gaussian~\cite{GVW:nat03,JKL+13,JCM21}, it is much more involved in the case of a discrete modulation. There, one needs to solve a semidefinite program whose dimension scales both with the constellation size and the dimension of the relevant Hilbert space -- infinite for CV protocols. Even if it is possible to truncate the Fock space to a relevant subspace~\cite{UHJ21}, this numerical approach quickly becomes untractable as soon as the constellation size exceeds 10. The main contribution of Ref.~\cite{denys21} is to provide an analytical formula for the covariance matrix, depending only on easily measurable quantities in the PM protocol, namely the variance of Bob's measurement result and two correlation coefficients between Alice and Bob's data. This will be analyzed further below.

\begin{figure}
\centering
\includegraphics[trim=3.5cm 0.8cm 1.2cm 1.5cm, clip, width=\columnwidth]{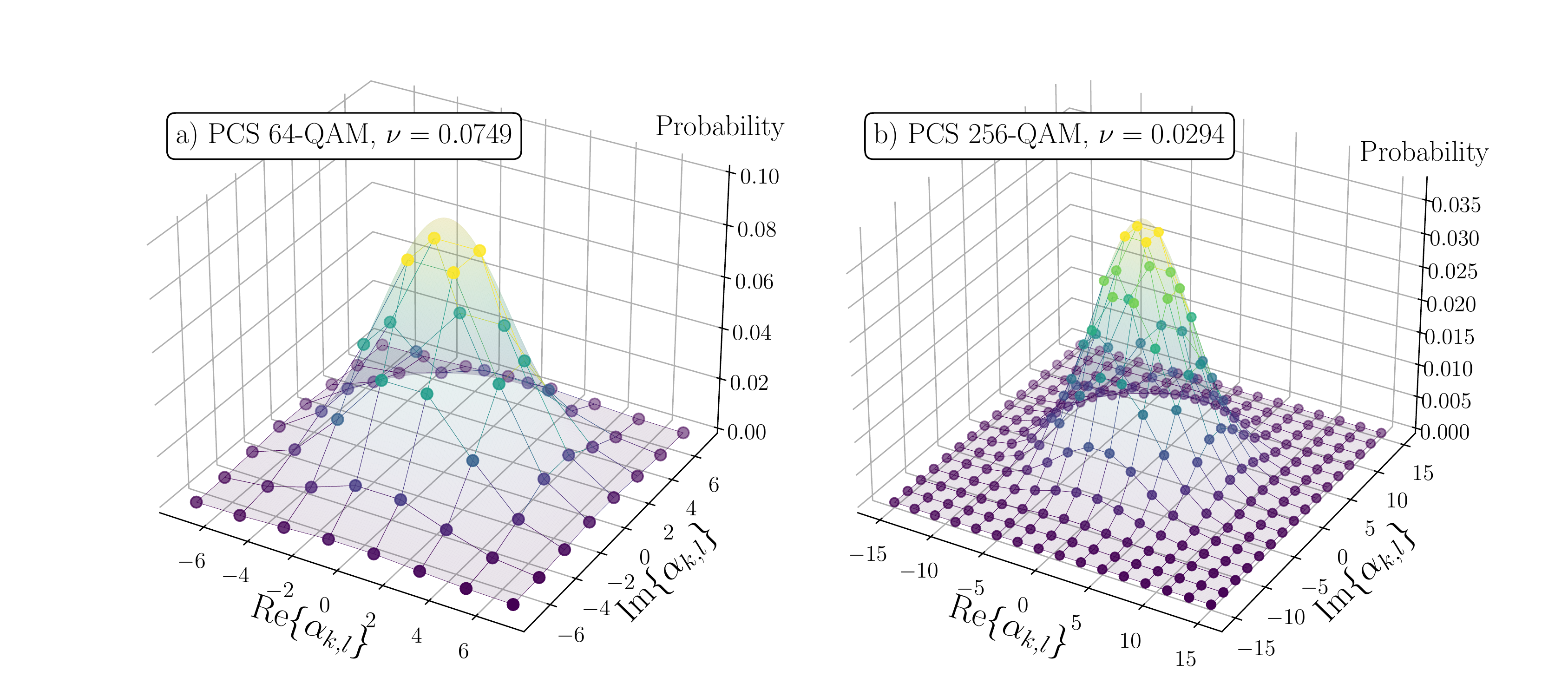}
\caption{Constellation probability distributions for (a) PCS 64-QAM with $\nu = 0.0749$, and (b) PCS 256-QAM with $\nu = 0.0294$. In both cases bottom units are $\sqrt{SNU}$ and $\alpha_0=2$ $\sqrt{SNU}$. Connecting lines and equivalent Gaussian distributions are depicted for clarity. The free parameter $\nu$ appears in the discretized Gaussian probability distribution describing the constellation, and its optimization is crucial for the maximization of the SKR.}
\label{fig:constellations}
\end{figure}

\paragraph{PCS QAM for CV-QKD.} The probabilistic constellation shaping with quadrature amplitude modulation (PCS QAM) is a standard modulation method \cite{proakis}, involving a discretized Gaussian probability distribution $\pi_{k,l}$  given by
\begin{align}
\alpha_{k,l} &= \alpha_0(k+il)\\
\pi_{k,l} &= \frac{\exp(-\nu|\alpha_{k,l}|^2)}{\sum_{k,l}\exp(-\nu|\alpha_{k,l}|^2)},
\end{align}
where $k+il$ are the points of a standard QAM constellation, and $\nu >0$ and $\alpha_0>0$ are free parameters such that $\sum_{k,l}\pi_{k,l}|\alpha_{k,l}|^2 = V_A/2$. 
Here, $V_A$ is the variance of Alice's modulation, measured in shot-noise units (SNU), \emph{i.e.}, such that the variance of the shot noise equals one. Since PCS QAM are commonly used in modern high-rate coherent optical transmission systems, very efficient digital signal processing techniques have been developed. Moreover, PCS QAM are good candidates for discrete modulation with near optimal SKR, thanks to their Gaussian-like distribution~\cite{OFC2021}. When using PCS QAM, it is crucial to optimize the free parameter $\nu$ to maximize the SKR. Using numerical calculation, we observed that the optimal value depends only on Alice's modulation variance $V_A$. In the following, both $V_A$ and $\nu$ are chosen to maximize the SKR for either 64-QAM or 256-QAM modulations, as displayed on Fig.~\ref{fig:constellations}.

\begin{figure}[h]
\centering
\includegraphics[width=\columnwidth]{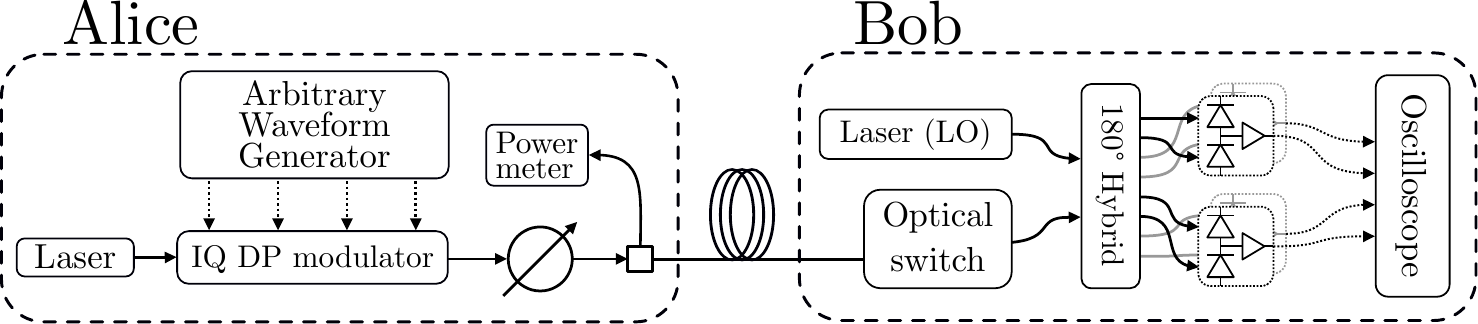}
\caption{Experimental setup. The setup only involves off-the-shelf, state-of-the-art telecom equipment. The 1550-nm laser has a 10 kHz nominal linewidth. The Arbitrary Waveform Generator feeds the dual-polarization in-phase-and-quadrature (DP-IQ) modulator with four 600 MBaud signals with Root Raised Cosine (RRC) pulse shape. An optical power meter and an attenuator at the output of Alice are used to monitor $V_A$. Bob uses a 180 degrees hybrid to interfere the signal with the local oscillator (LO) and a set of four amplified balanced photodetectors, whose outputs are sampled using a real-time oscilloscope and processed by offline digital signal processing. At the input of Bob, a microelectromechanical optical switch is used to periodically turn off the signal to perform shot noise measurements for noise calibration.}
\label{fig:setup}
\end{figure}

\paragraph{Experimental implementation.} The main idea behind the development of the experimental system in our work is to use only commercially available, latest generation telecom equipment in order to provide a convenient cost-efficient solution. Important requirements that we sought for were high resolution, low noise and a bandwidth of at least 1 GHz. The setup is shown in Fig.~\ref{fig:setup}. Alice generates coherent states using a 1550 nm tunable laser source with nominal 10 kHz linewidth (Pure Photonics). A dual polarization (DP) in-phase-and-quadrature (IQ) modulator (Fujitsu) is used to modulate the phase and amplitude of the laser beam. The analog inputs of the DP-IQ modulator are fed with the output of an Arbitrary Waveform Generator (AWG) with 5 GS/s sampling rate and 14 bits nominal resolution. The AWG outputs four 600 MBaud signals with Root Raised Cosine (RRC) pulse shape~\cite{proakis}. At the output of Alice's lab, an optical power meter and an optical attenuator are used to monitor $V_A$.
Bob uses a 180 degrees hybrid to interfere the signal with the phase reference (or local oscillator, LO), which is generated with a laser identical to Alice's. Four amplified balanced photodetectors convert the received optical signal to an analog electronic signal, which is then sampled using a 1 GHz real-time oscilloscope with 5 GS/s sampling rate and 12 bits nominal resolution. The sampled waveforms are stored for offline digital signal processing (DSP, see below).
In the present experiment, the memory and writing speed of the oscilloscope impose to perform noise calibration and parameter estimation one acquisition at a time, but in a full-scale implementation the oscilloscope and offline DSP would be replaced by a continuously running receiver with real-time DSP.

\begin{figure}
\centering
\includegraphics[trim=3cm 1cm 4cm 0cm, width=0.9\columnwidth]{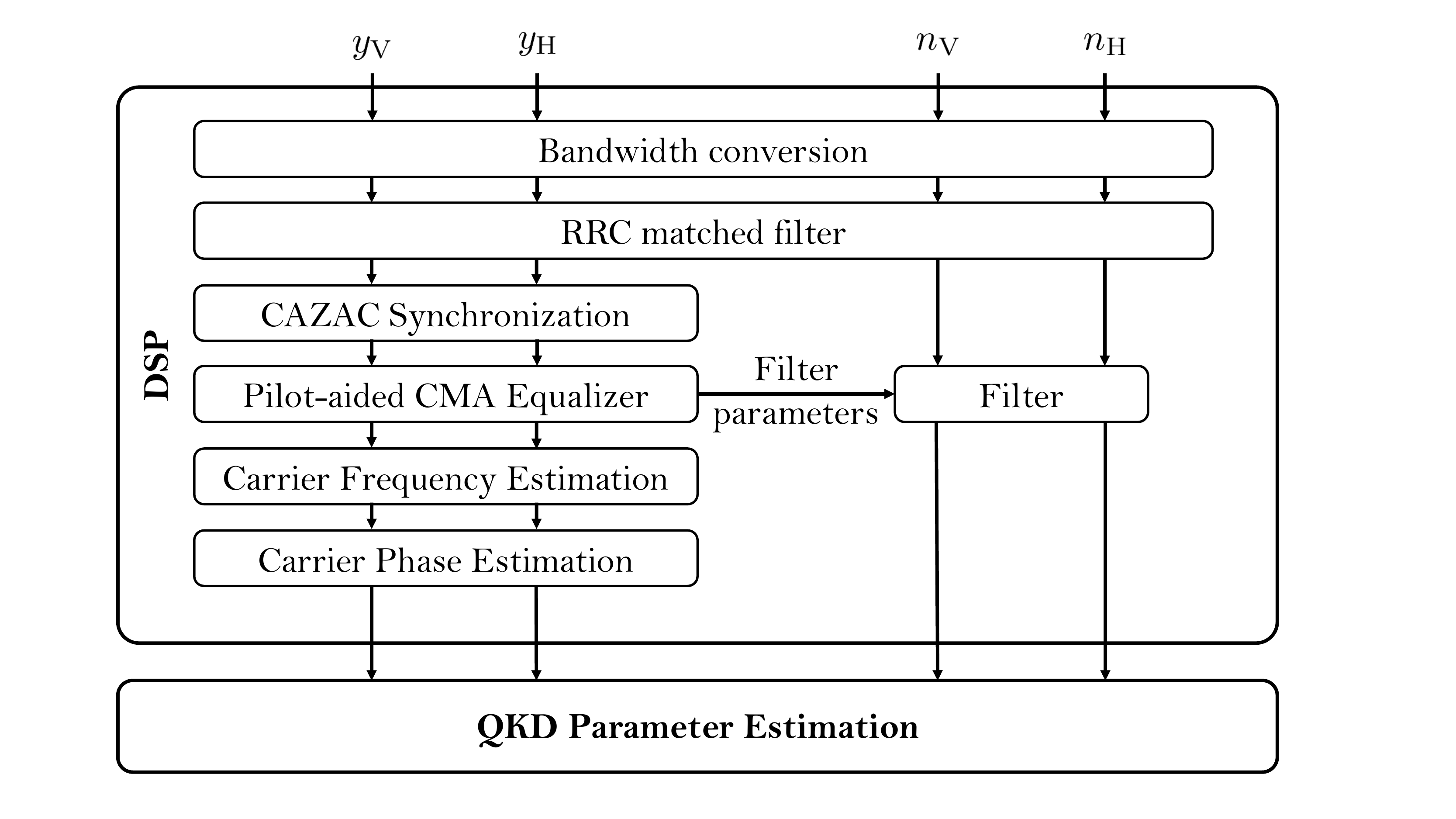}
\caption[Digital Signal Processing]{Bob's digital signal processing building blocks. DSP consists in a combination of digital filter matching the pulse shape of input symbols $y_H, y_V$, auto-correlation for retrieving the time-multiplexed pilots used in our experiments, a pilot-aided adaptive equalizer technique, and finally carrier frequency and phase estimation algorithms. It is then possible, by using the transmitted data together with noise calibration data $n_H, n_V$ that have undergone the same processing, to estimate the secret key rate.}
\label{fig:exp_dsp_blocks}
\end{figure}

\paragraph{Digital signal processing.} The implementation of DSP suitable for CV-QKD is one of the most important practical challenges of this work. The main building blocks are shown in Fig.~\ref{fig:exp_dsp_blocks}. The algorithm inputs four sampled waveforms $y_1(k)$, $y_2(k)$, $y_3(k)$, $y_4(k)$, with an average number of samples per transmitted symbol $\bar{n}_{\mathrm{sps}}= 8.3$ (calculated by dividing the  $5~\mathrm{GS/s}$ sampling rate with the $600~\mathrm{Mbaud}$ symbol rate). The waveforms are then assembled into two complex waveforms $y_H(k) = y_1(k) + j y_2(k)$ and $y_V(k) = y_3(k) + j y_4(k)$.
If the signal is single-side band (see details in Methods), it is converted into a baseband signal by a digital frequency shift, and a digital filter matching the pulse shape is applied; a root raised cosine (RRC) filter in our case.
Then, we use a constant amplitude zero autocorrelation waveform (CAZAC) sequence~\cite{milewski1983} to compute the auto-correlation on the signal in order to retrieve the beginning of the time-multiplexed pilot sequence used in our implementation.
The next steps are to correct linear impairments using a pilot-aided CMA adaptive equalizer~\cite{FMZ+:oe10} (see details in Methods), and to apply carrier frequency and carrier phase estimation algorithms. Finally, using the noise calibration symbols, denoted as $n_H$ and $n_V$ in Fig.~\ref{fig:exp_dsp_blocks}, which undergo the same DSP operations, QKD parameters are estimated to compute the achievable secret key rate.

These algorithms are obviously unable to perfectly correct channel impairments, and the DSP imperfections may result in apparent channel excess noise. Therefore it is crucial to optimize the various DSP parameters to minimize excess noise, ideally for each individual run of the experiment producing a block of data. In this work, the optimization procedure has been performed offline, after signal acquisition, and is described in Methods.

\paragraph{Noise calibration measurements.} Most of the CV-QKD parameters are expressed in SNU. However, Bob effectively measures samples $U$ of an electrical tension expressed in volts; see Methods for a description of the required calibration procedure.
We note that the LO intensity and thus the shot noise may vary during the experiment, making it necessary to periodically reiterate the procedure of recording shot noise samples as often as possible. For this purpose, Bob's setup includes an optical switch used to turn on and off the signal light coming from Alice. 
This procedure is repeated once every minute. Finally, the normalized value $V_B$ of Bob's variance can be written as
\begin{equation}
V_B = 1+  \eta T V_A/2  + V_{\mathrm{el}} + \xi_B,
\end{equation}
where $T$ is the channel transmission efficiency, and  $\xi_B$ is the excess noise measured at Bob's site, to be evaluated by Alice and Bob. The quantum efficiency and electronic noise of Bob's detectors, which in our experiment take the values $\eta = 0.65$ and $V_{\mathrm{el}}= 0.1$, respectively, are supposed here to be known to the legitimate users and cannot be modified by Eve.

\paragraph{Non-Gaussian attacks.} Recall that in our protocol, which follows the security proof of Ref.~\cite{denys21}, Alice and Bob should not in fact evaluate $T$ and $\xi_B$ from the data, but rather three parameters, denoted as $c_1$, $c_2$ and $n_B$. Under the assumption of a Gaussian channel these parameters are simply related to $T$ and $\xi_B$, and to the parameters defining the constellation~\cite{FRthesis}. However, the Gaussian channel assumption is not justified for an arbitrary attack by Eve on a discrete modulation, and $c_1$, $c_2$ and $n_B$ must be evaluated directly. As a consequence, the SKR, related to the Holevo quantity, is a function $f(c_1, c_2, n_B)$, instead of the usual $g(T,\xi_B)$; see Ref.~\cite{FRthesis}.

Under our experimental conditions, we found the effective channel to be very well described by a Gaussian model, and we observe $f(\hat c_1, \hat c_2, \hat n_B) \simeq g(\hat T,\hat \xi_B)$. However, the direct evaluation of these formulas with the estimators ignores finite-size effects.
In order to take them into account, we evaluate the formulas with worst-case estimators~\cite{denys21}, \emph{i.e.}, we rather compute $f(\hat c_1^\text{min}, \hat c_2^\text{min}, \hat n_B^\text{max})$, which is less favorable than $g(\hat T^\text{min}, \hat \xi_B^\text{max})$ for a Gaussian channel.
This is the procedure followed to obtain the results provided in Table~\ref{tab:new_results_c1_c2_nB}, which correspond to a rigorous implementation of the protocol with the security proof for a discrete modulation~\cite{FRthesis}.

\begin{table}
\centering
\begin{tabular}{|c|c|c|c|c|c|}
\hline
Fiber & Modulation &  $\nu$ & $V_A$[\text{SNU}] & $\xi_B$[\text{mSNU}] & SKR[\text{Mbps}] \\ \hline\hline
\multirow{2}{*}{\shortstack{9.5 km \\ SMF-28}} & 64-QAM & 0.0688 & 5.32 & 0.197 & 60.2 \\ \cline{2-6}
& 256-QAM &  0.0362 & 7.11 & 0.132 & 91.8 \\ \hline \hline
\multirow{2}{*}{\shortstack{25 km \\ EX3000}} & 64-QAM &   0.0460 & 4.20 & 1.170 &   0.0 \\ \cline{2-6}
 & 256-QAM &  0.0380 & 6.53 & 0.900 & 24.0 \\ \hline
\end{tabular}
\caption{Modulation variance $V_A$ (in \text{SNU}), indicative excess noise $\xi_B$ (in \text{mSNU}) and SKR calculated using the security proof of Ref.~\cite{denys21} including finite-size effects (in \text{Mbps}), for PCS 64-QAM and PCS 256-QAM, during 1 hour of experiment, with 9.5 km of SMF-28 and 25 km of EX3000 fiber. The block size is $N=5\times 10^6$.}
\label{tab:new_results_c1_c2_nB}
\end{table}

\paragraph{Results.} Our experiment was performed with either 9.5 km of SMF-28 or 25 km of EX3000 fiber. The 25 km fiber link has a total loss of 4.3~dB. In each case the most critical DSP parameters are optimized to minimize the excess noise. In the present implementation the system operates with acquisitions of length 20 ms from which, after processing, $N=5\times 10^6$  QKD symbols are used for parameter estimation. Finally, the DSP optimization process is performed on a subset of 12 acquisitions.

\begin{figure}
\centering
\includegraphics[trim=1cm 1cm 1cm 0cm, width=\columnwidth]{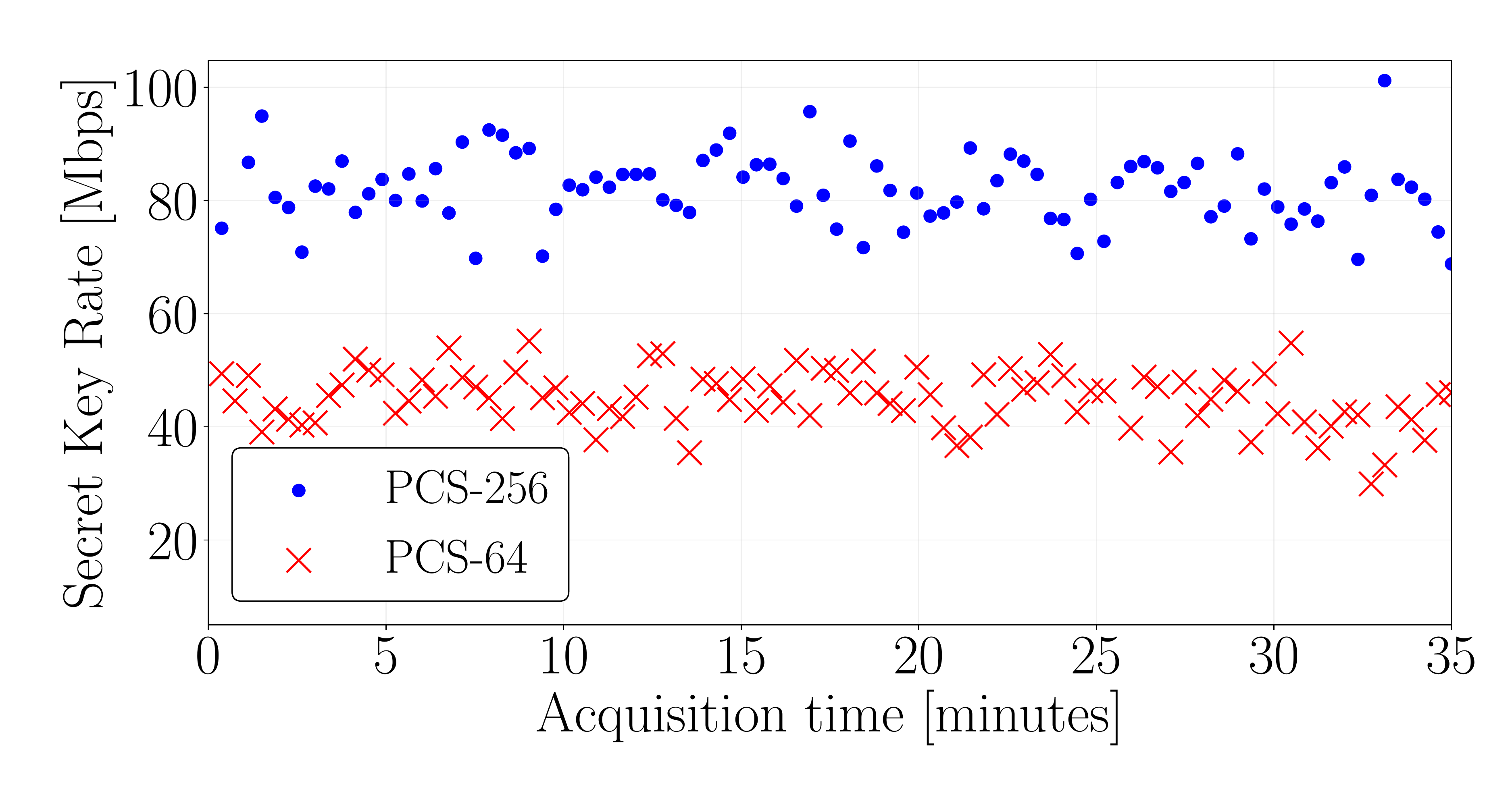}
\caption{Estimated  secret key rate for each block of acquired data, plotted as a function of the acquisition time, for PCS 64 and 256-QAM, with 9.5 km SMF-28 link.
\label{fig:new_xi_skr_10km}}
\end{figure}

Figure~\ref{fig:new_xi_skr_10km}  shows the estimated SKR for the 9.5 km SMF-28  fiber, for PCS 64 and 256-QAM. The estimation is based on the proof for an arbitrary modulation protocol~\cite{denys21}, assuming $\beta=0.95$~\cite{JKL:pra11}, and using worst-case estimators with $N=5\times 10^6$ and security parameter $\epsilon = 10^{-10}$, following Refs.~\cite{SR:prl08,LGG:pra10,JKDL:pra12}. Table~\ref{tab:new_results_c1_c2_nB} summarizes the results with modulation variance $V_A$ values (in SNU), excess noise $\xi_B$ values (in SNU), which are included as an indication of system performance, and SKR (in Mbps) calculated following the aforementioned procedure.

\begin{figure}
\centering
\includegraphics[trim=1cm 1cm 2cm 0cm, width=\columnwidth]{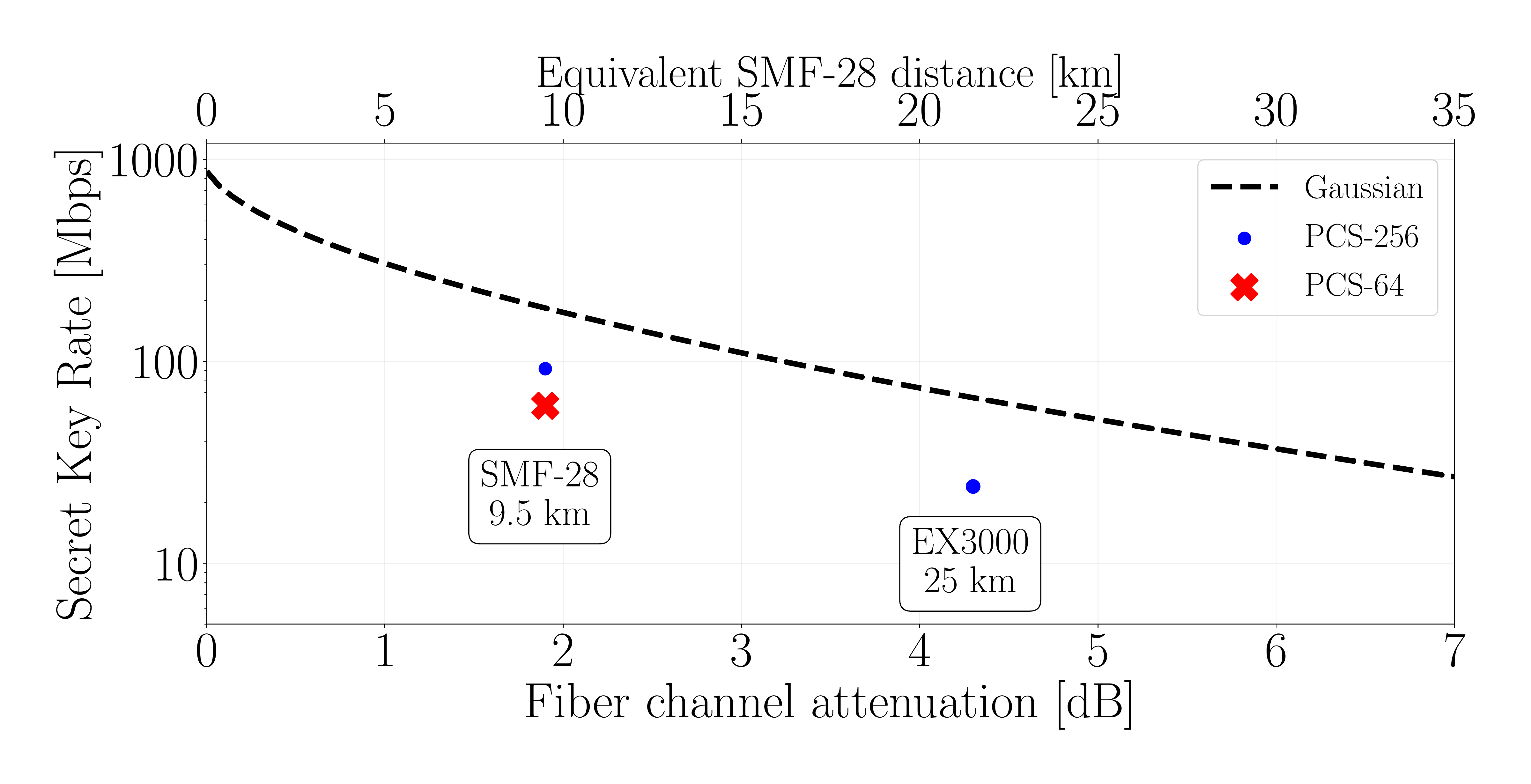}
\caption{Experimental results of secret key rate as a function of the channel attenuation and distance considering finite-size effects and neglecting post-processing times. Two modulation formats (PCS-64 and PCS-256) and two fibers have been used in this experiment; a 9.5 km standard single mode fiber (SMF-28) with attenuation coefficient 0.2 dB/km and a 25 km EX3000 fiber with attenuation coefficient 0.172 dB/km. PCS-64 modulation at 25 km does not yield a positive key rate. The expected SKR of a setup with Gaussian modulation in the asymptotic regime is plotted for comparison, assuming the same repetition rate $R=600$ MBaud, $\xi_B = 0.5$ mSNU, and Alice using the optimal $V_A$. The block size is $N=5\times 10^6$.}
\label{distance}
\end{figure}

We can achieve a secret key rate of $\sim$ 92~Mbps over 9.5~km and 24~Mbps over 25~km, using PCS 256-QAM format, averaged over 100 transmission blocks of $N=5\times 10^6$ QKD symbols. PCS 64-QAM  gives lower performance, as theoretically expected.  The expected behavior as a function of distance is shown in Fig.~\ref{distance}. By comparison with the current state of the art~\cite{pir2020,xu2020,ZCP+:prl20,JCM21,PWS:ol22}, these results confirm the high performance reached by our system by adopting techniques from standard optical communication and following the security proof for discrete modulation, including finite-size effects.

\paragraph{Conclusion.} The laboratory experiment presented in this work opens interesting avenues towards faster and more flexible implementations of CV-QKD, within the standard environment of high bit rate coherent telecommunications. It leverages in particular industry-grade digital signal processing techniques that have been minimally modified for the CV-QKD implementation. To take full advantage of these improvements, it would be necessary to also improve the speed of data post-processing, which should be facilitated by the use of discrete constellations.

\vspace{-5mm}
\section{Methods}
\vspace{-3mm}

\paragraph{Pilot amplitude and rate.} To correctly retrieve the low signal-to-noise ratio QKD symbols, the DSP relies on QPSK (that is 4-QAM) pilot symbols with a higher power, which needs to be optimized before signal acquisition. This is done by acquiring QKD signals with various values of the pilot amplitude, and applying the DSP to estimate the excess noise.
Using such experimental tests, the pilot over QKD symbol power ratio was adjusted to 14 dB.
The same optimization should be performed for the pilot rate. Contrary to pilot amplitude, the criterion to optimize the pilot rate is not the excess noise. In fact, if an increase of the pilot rate decreases the excess noise, it also decreases the rate of QKD symbols. Hence, we need to optimize directly the SKR.
Using again an experimental optimization, we fixed the pilot rate to 4 pilots over 8 symbols, \emph{i.e.}, half of the transmitted symbols are actually pilots.

\begin{figure}
\centering
\includegraphics[trim=0.5cm 0cm 0.5cm 0cm, width=0.8\columnwidth]{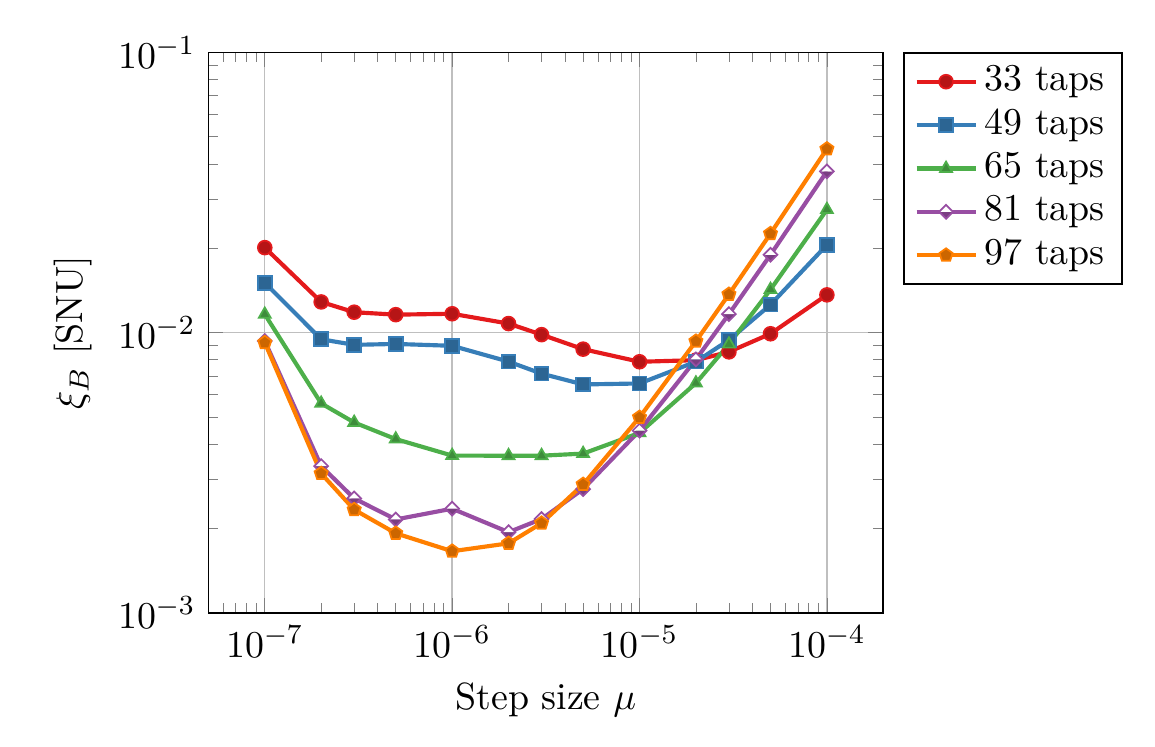}
\caption{Excess noise $\xi_B$ vs. step size $\mu$ and number of taps of adaptive equalizer, averaged over 12 acquisitions of PCS 256-QAM signal and 25 km of EX3000 fiber.}
\label{fig:taps}
\end{figure}

\paragraph{Adaptive equalizer.} For each experiment, we want to find the DSP parameters that minimize the excess noise. Since the DSP is performed offline, we can do a brute force optimization for the most relevant parameters, on a few acquisitions. To start with, we jointly optimize two parameters of the adaptive equalizer for polarization demultiplexing~\cite{Kikuchi16}: $n_{\mathrm{taps}}$, number of taps, and $\mu$, the step size. For each couple $(n_{\mathrm{taps}},\mu)$ under test, the DSP is applied to 12 different acquisitions.
Figure~\ref{fig:taps} shows the average excess noise for all the tested $(n_{\mathrm{taps}},\mu)$, for experimental PCS 256-QAM data obtained in conditions slightly different than those presented in the main text. We observe that the lowest values of excess noise are achieved with 97 taps and a step size $\mu$ of $10^{-6}$.

\paragraph{Signal conditioning.} We observed the presence of low frequency components of the excess noise, below 20 MHz, that we attribute to cutoff frequencies of the hardware as well as additive noise stemming from the electrical line. To avoid these perturbations, the outputs of the AWG are digitally upshifted such that the signal has no frequency component in the noisy region. In particular, the 600 MBaud signal with RRC pulse shape and roll-off factor 0.4, corresponding to a bandwidth of 840 MHz, is upshifted by 500 MHz such that the useful bandwidth extends from 80 MHz to 920 MHz.
The baudrate and roll-off factor have to be jointly optimized to minimize the excess noise. Furthermore, as noted above, the ratio of the QPSK pilots power relatively to the QAM symbols power has to be optimized to minimize the excess noise.

\paragraph{Noise calibration.} Since Bob effectively measures samples $U$ of an electrical tension expressed in volts, and obtains variances $\mathrm{Var}(U)$ in {$V^2$}, he needs to estimate the quantity $N_0$, namely the variance of the shot noise expressed in {$V^2$}. When disconnecting the signal input of the receiver, the output of the receiver is the sum of the shot noise and the electronic noise; therefore Bob can measure
$\mathrm{Var}(U) = N_0(1+V_{\mathrm{el}})$, where $V_{\mathrm{el}}$ is the variance of the receiver's electronic noise in SNU. Then, disconnecting the LO input, Bob measures only the electronic noise, $\mathrm{Var}(U') = N_0V_{\mathrm{el}}$ and $N_0 = \mathrm{Var}(U) - \mathrm{Var}(U')$.

This procedure gives four different values $N_0^{(1)}$, $N_0^{(2)}$, $N_0^{(3)}$, and $N_0^{(4)}$, one for each channel of the oscilloscope. In practice, the samples measured on a channel are a mixture of the quadratures of the coherent states sent by Alice, that are recovered only after the DSP. This comes from several channel impairments such as polarization rotation or carrier phase noise. As a consequence, if the $N_0^{(i)}$ are not all equal, they do not correspond to the variances of the shot noise on the quadratures effectively transmitted by Alice. To tackle this issue, we apply to the shot noise samples the same DSP correction as to the signal itself, and estimate the variances afterwards. In other words, the DSP operations applied to the signal samples are simultaneously applied to the noise samples.

\paragraph{Signal averaging.} Our use of a worst-case estimator is justified if the fluctuations observed on the parameters are of a statistical nature. Given that all $5 \times 10^6$ data points within a data block are very close in time (total acquisition time 20 ms), the population variance can be considered sufficiently close to the theoretical variance to assume that fluctuations on the excess noise measurement are essentially of statistical nature. Therefore, the use of the worst-case estimator for the excess noise can be considered acceptable to take into account finite-size effects on the security of the protocol, although a more rigorous theoretical treatment of finite-size issues remains desirable.

\section{Acknowledgments}
This research was supported by the E.C. projects CiViQ and OpenQKD, with a contribution by the Paris Region project ParisRegionQCI. F.R. was supported by a CIFRE PhD grant.


\begin{thebibliography}{10}
\expandafter\ifx\csname url\endcsname\relax
  \def\url#1{\texttt{#1}}\fi
\expandafter\ifx\csname urlprefix\endcsname\relax\def\urlprefix{URL }\fi
\providecommand{\bibinfo}[2]{#2}
\providecommand{\eprint}[2][]{\url{#2}}

\bibitem{SBC:rmp09}
\bibinfo{author}{Scarani, V.} \emph{et~al.}
\newblock \bibinfo{title}{The security of practical quantum key distribution}.
\newblock \emph{\bibinfo{journal}{Rev. Mod. Phys.}}
  \textbf{\bibinfo{volume}{81}}, \bibinfo{pages}{1301} (\bibinfo{year}{2009}).

\bibitem{WPG:rmp12}
\bibinfo{author}{Weedbrook, C.} \emph{et~al.}
\newblock \bibinfo{title}{Gaussian quantum information}.
\newblock \emph{\bibinfo{journal}{Rev. Mod. Phys.}}
  \textbf{\bibinfo{volume}{84}}, \bibinfo{pages}{621} (\bibinfo{year}{2012}).

\bibitem{DL15}
\bibinfo{author}{Diamanti, E.} \& \bibinfo{author}{Leverrier, A.}
\newblock \bibinfo{title}{Distributing secret keys with quantum continuous
  variables: Principle, security and implementations}.
\newblock \emph{\bibinfo{journal}{Entropy}} \textbf{\bibinfo{volume}{6072}},
  \bibinfo{pages}{17} (\bibinfo{year}{2015}).

\bibitem{JKL+13}
\bibinfo{author}{Jouguet, P.}, \bibinfo{author}{Kunz-Jacques, S.},
  \bibinfo{author}{Leverrier, A.}, \bibinfo{author}{Grangier, P.} \&
  \bibinfo{author}{Diamanti, E.}
\newblock \bibinfo{title}{Experimental demonstration of long-distance
  continuous-variable quantum key distribution}.
\newblock \emph{\bibinfo{journal}{Nature Photon.}}
  \textbf{\bibinfo{volume}{7}}, \bibinfo{pages}{378} (\bibinfo{year}{2013}).

\bibitem{ZCP+:prl20}
\bibinfo{author}{Zhang, Y.} \emph{et~al.}
\newblock \bibinfo{title}{Long-distance continuous-variable quantum key
  distribution over 202.81 km of fiber}.
\newblock \emph{\bibinfo{journal}{Phys. Rev. Lett.}}
  \textbf{\bibinfo{volume}{125}}, \bibinfo{pages}{010502}
  (\bibinfo{year}{2020}).

\bibitem{JCM21}
\bibinfo{author}{Jain, N.} \emph{et~al.}
\newblock \bibinfo{title}{Practical continuous-variable quantum key
  distribution with composable security}.
\newblock \emph{\bibinfo{journal}{arXiv:2110.09262 [quant-ph]}}
  (\bibinfo{year}{2021}).

\bibitem{denys21}
\bibinfo{author}{Denys, A.}, \bibinfo{author}{Brown, P.} \&
  \bibinfo{author}{Leverrier, A.}
\newblock \bibinfo{title}{Explicit asymptotic secret key rate of
  continuous-variable quantum key distribution with an arbitrary modulation}.
\newblock \emph{\bibinfo{journal}{Quantum}} \textbf{\bibinfo{volume}{5}},
  \bibinfo{pages}{540} (\bibinfo{year}{2021}).

\bibitem{DLQY:npjQI16}
\bibinfo{author}{Diamanti, E.}, \bibinfo{author}{Lo, H.-K.},
  \bibinfo{author}{Qi, B.} \& \bibinfo{author}{Yuan, Z.}
\newblock \bibinfo{title}{Practical challenges in quantum key distribution}.
\newblock \emph{\bibinfo{journal}{npj Quant. Inf.}}
  \textbf{\bibinfo{volume}{2}}, \bibinfo{pages}{16025} (\bibinfo{year}{2016}).

\bibitem{pir2020}
\bibinfo{author}{Pirandola, S.} \emph{et~al.}
\newblock \bibinfo{title}{Advances in quantum cryptography}.
\newblock \emph{\bibinfo{journal}{Adv. Opt. Photonics}}
  \textbf{\bibinfo{volume}{12}}, \bibinfo{pages}{1012} (\bibinfo{year}{2020}).

\bibitem{xu2020}
\bibinfo{author}{Xu, F.}, \bibinfo{author}{Ma, X.}, \bibinfo{author}{Zhang,
  Q.}, \bibinfo{author}{Lo, H.-K.} \& \bibinfo{author}{Pan, J.-W.}
\newblock \bibinfo{title}{Secure quantum key distribution with realistic
  devices}.
\newblock \emph{\bibinfo{journal}{Rev. Mod. Phys.}}
  \textbf{\bibinfo{volume}{92}}, \bibinfo{pages}{025002}
  (\bibinfo{year}{2020}).

\bibitem{GVW:nat03}
\bibinfo{author}{Grosshans, F.} \emph{et~al.}
\newblock \bibinfo{title}{Quantum key distribution using gaussian-modulated
  coherent states}.
\newblock \emph{\bibinfo{journal}{Nature}} \textbf{\bibinfo{volume}{421}},
  \bibinfo{pages}{238} (\bibinfo{year}{2003}).

\bibitem{LRH+:pra06}
\bibinfo{author}{Lorenz, S.} \emph{et~al.}
\newblock \bibinfo{title}{Witnessing effective entanglement in a continuous
  variable prepare-and-measure setup and application to a quantum key
  distribution scheme using postselection}.
\newblock \emph{\bibinfo{journal}{Phys. Rev. A}} \textbf{\bibinfo{volume}{74}},
  \bibinfo{pages}{042326} (\bibinfo{year}{2006}).

\bibitem{LG09}
\bibinfo{author}{Leverrier, A.} \& \bibinfo{author}{Grangier, P.}
\newblock \bibinfo{title}{Unconditional security proof of long-distance
  continuous-variable quantum key distribution with discrete modulation}.
\newblock \emph{\bibinfo{journal}{Phys. Rev. Lett.}}
  \textbf{\bibinfo{volume}{102}}, \bibinfo{pages}{180504}
  (\bibinfo{year}{2009}).

\bibitem{LG11}
\bibinfo{author}{Leverrier, A.} \& \bibinfo{author}{Grangier, P.}
\newblock \bibinfo{title}{Continuous-variable quantum-key-distribution
  protocols with a non-gaussian modulation}.
\newblock \emph{\bibinfo{journal}{Phys. Rev. A}} \textbf{\bibinfo{volume}{83}},
  \bibinfo{pages}{042312} (\bibinfo{year}{2011}).

\bibitem{KGW21}
\bibinfo{author}{Kaur, E.}, \bibinfo{author}{Guha, S.} \&
  \bibinfo{author}{Wilde, M.~M.}
\newblock \bibinfo{title}{Asymptotic security of discrete-modulation protocols
  for continuous-variable quantum key distribution}.
\newblock \emph{\bibinfo{journal}{Phys. Rev. A}}
  \textbf{\bibinfo{volume}{103}}, \bibinfo{pages}{012412}
  (\bibinfo{year}{2021}).

\bibitem{GGDL19}
\bibinfo{author}{Ghorai, S.}, \bibinfo{author}{Grangier, P.},
  \bibinfo{author}{Diamanti, E.} \& \bibinfo{author}{Leverrier, A.}
\newblock \bibinfo{title}{Asymptotic security of continuous-variable quantum
  key distribution with a discrete modulation}.
\newblock \emph{\bibinfo{journal}{Phys. Rev. X}} \textbf{\bibinfo{volume}{9}},
  \bibinfo{pages}{021059} (\bibinfo{year}{2019}).

\bibitem{LUL19}
\bibinfo{author}{Lin, J.}, \bibinfo{author}{Upadhyaya, T.} \&
  \bibinfo{author}{L\"utkenhaus, N.}
\newblock \bibinfo{title}{Asymptotic security analysis of discrete-modulated
  continuous-variable quantum key distribution}.
\newblock \emph{\bibinfo{journal}{Phys. Rev. X}} \textbf{\bibinfo{volume}{9}},
  \bibinfo{pages}{041064} (\bibinfo{year}{2019}).

\bibitem{LL20}
\bibinfo{author}{Lin, J.} \& \bibinfo{author}{L\"utkenhaus, N.}
\newblock \bibinfo{title}{Trusted detector noise analysis for discrete
  modulation schemes of continuous-variable quantum key distribution}.
\newblock \emph{\bibinfo{journal}{Phys. Rev. Appl.}}
  \textbf{\bibinfo{volume}{14}}, \bibinfo{pages}{064030}
  (\bibinfo{year}{2020}).

\bibitem{MMS21}
\bibinfo{author}{Matsuura, T.}, \bibinfo{author}{Maeda, K.},
  \bibinfo{author}{Sasaki, T.} \& \bibinfo{author}{Koashi, M.}
\newblock \bibinfo{title}{Finite-size security of continuous-variable quantum
  key distribution with digital signal processing}.
\newblock \emph{\bibinfo{journal}{Nature Comm.}} \textbf{\bibinfo{volume}{12}},
  \bibinfo{pages}{252} (\bibinfo{year}{2021}).

\bibitem{HIM+:qst17}
\bibinfo{author}{Hirano, T.} \emph{et~al.}
\newblock \bibinfo{title}{Implementation of continuous-variable quantum key
  distribution with discrete modulation}.
\newblock \emph{\bibinfo{journal}{Quant. Sci. Tech.}}
  \textbf{\bibinfo{volume}{2}}, \bibinfo{pages}{024010} (\bibinfo{year}{2017}).

\bibitem{WLL+:arXiv21}
\bibinfo{author}{Wang, P.}, \bibinfo{author}{Liu, J.}, \bibinfo{author}{Lu,
  Z.}, \bibinfo{author}{Wang, X.} \& \bibinfo{author}{Li, Y.}
\newblock \bibinfo{title}{Discrete-modulation continuous-variable quantum key
  distribution with high key rate}.
\newblock \emph{\bibinfo{journal}{arXiv:2112.00214 [quant-ph]}}
  (\bibinfo{year}{2021}).

\bibitem{GFR17}
\bibinfo{author}{Ghazisaeidi, A.} \emph{et~al.}
\newblock \bibinfo{title}{Advanced c+l-band transoceanic transmission systems
  based on probabilistically shaped pdm-64qam}.
\newblock \emph{\bibinfo{journal}{J. Lightwave Tech.}}
  \textbf{\bibinfo{volume}{35}}, \bibinfo{pages}{1291} (\bibinfo{year}{2017}).

\bibitem{ECOC2021}
\bibinfo{author}{Roumestan, F.} \emph{et~al.}
\newblock \bibinfo{title}{High-rate continuous variable quantum key
  distribution based on probabilistically shaped 64 and 256-qam}.
\newblock In \emph{\bibinfo{booktitle}{European Conference on Optical
  Communication (ECOC)}} (\bibinfo{address}{Bordeaux, France},
  \bibinfo{year}{2021}).
\newblock \bibinfo{note}{Doi:10.1109/ECOC52684.2021.9606013t}.

\bibitem{PWS:ol22}
\bibinfo{author}{Pan, Y.} \emph{et~al.}
\newblock \bibinfo{title}{Experimental demonstration of high-rate
  discrete-modulated continuous-variable quantum key distribution system}.
\newblock \emph{\bibinfo{journal}{Opt. Lett.}} \textbf{\bibinfo{volume}{47}},
  \bibinfo{pages}{3307} (\bibinfo{year}{2022}).

\bibitem{WGC06}
\bibinfo{author}{Wolf, M.~M.}, \bibinfo{author}{Giedke, G.} \&
  \bibinfo{author}{Cirac, J.~I.}
\newblock \bibinfo{title}{Extremality of gaussian quantum states}.
\newblock \emph{\bibinfo{journal}{Phys. Rev. Lett.}}
  \textbf{\bibinfo{volume}{96}}, \bibinfo{pages}{080502}
  (\bibinfo{year}{2006}).

\bibitem{GC:prl06}
\bibinfo{author}{Garc\'{\i}a-Patr\'{o}n, R.} \& \bibinfo{author}{Cerf, N.~J.}
\newblock \bibinfo{title}{Unconditional optimality of gaussian attacks against
  continuous-variable quantum key distribution}.
\newblock \emph{\bibinfo{journal}{Phys. Rev. Lett.}}
  \textbf{\bibinfo{volume}{97}}, \bibinfo{pages}{190503}
  (\bibinfo{year}{2006}).

\bibitem{NGA:prl06}
\bibinfo{author}{Navascu\'{e}s, M.}, \bibinfo{author}{Grosshans, F.} \&
  \bibinfo{author}{Ac\'{\i}n, A.}
\newblock \bibinfo{title}{Optimality of gaussian attacks in continuous-variable
  quantum cryptography}.
\newblock \emph{\bibinfo{journal}{Phys. Rev. Lett.}}
  \textbf{\bibinfo{volume}{97}}, \bibinfo{pages}{190502}
  (\bibinfo{year}{2006}).

\bibitem{UHJ21}
\bibinfo{author}{Upadhyaya, T.}, \bibinfo{author}{van Himbeeck, T.},
  \bibinfo{author}{Lin, J.} \& \bibinfo{author}{L\"utkenhaus, N.}
\newblock \bibinfo{title}{Dimension reduction in quantum key distribution for
  continuous- and discrete-variable protocols}.
\newblock \emph{\bibinfo{journal}{PRX Quantum}} \textbf{\bibinfo{volume}{2}},
  \bibinfo{pages}{020325} (\bibinfo{year}{2021}).

\bibitem{proakis}
\bibinfo{author}{Proakis, J.} \& \bibinfo{author}{Salehi, M.}
\newblock \emph{\bibinfo{title}{Digital Communications}}
  (\bibinfo{publisher}{McGraw-Hill Higher Education}, \bibinfo{year}{2007}).

\bibitem{OFC2021}
\bibinfo{author}{Roumestan, F.} \emph{et~al.}
\newblock \bibinfo{title}{Demonstration of probabilistic constellation shaping
  for continuous variable quantum key distribution}.
\newblock In \emph{\bibinfo{booktitle}{Optical Fiber Communication (OFC)}}
  (\bibinfo{address}{Washington, United States}, \bibinfo{year}{2021}).
\newblock \bibinfo{note}{Paper F4E.1}.

\bibitem{milewski1983}
\bibinfo{author}{Milewski, A.}
\newblock \bibinfo{title}{Periodic sequences with optimal properties for
  channel estimation and fast start-up equalization}.
\newblock \emph{\bibinfo{journal}{IBM Journal of Research and Development}}
  \textbf{\bibinfo{volume}{27}}, \bibinfo{pages}{426} (\bibinfo{year}{1983}).

\bibitem{FMZ+:oe10}
\bibinfo{author}{Faruk, M.~S.}, \bibinfo{author}{Mori, Y.},
  \bibinfo{author}{Zhang, C.}, \bibinfo{author}{Igarashi, K.} \&
  \bibinfo{author}{Kikuchi, K.}
\newblock \bibinfo{title}{Multiimpairment monitoring from adaptive
  finite-impulse-response filters in a digital coherent receiver,}.
\newblock \emph{\bibinfo{journal}{Opt. Express}} \textbf{\bibinfo{volume}{18}},
  \bibinfo{pages}{26929} (\bibinfo{year}{2010}).

\bibitem{FRthesis}
\bibinfo{author}{Roumestan, F.}
\newblock \emph{\bibinfo{title}{Advanced signal processing techniques for
  optical fiber continuous-variable quantum key distribution systems}}.
\newblock Ph.D. thesis, \bibinfo{school}{Sorbonne Université}
  (\bibinfo{year}{2022}).
\newblock \bibinfo{note}{Available online at
  https://tel.archives-ouvertes.fr/tel-03707442v1}.

\bibitem{JKL:pra11}
\bibinfo{author}{Jouguet, P.}, \bibinfo{author}{Kunz-Jacques, S.} \&
  \bibinfo{author}{Leverrier, A.}
\newblock \bibinfo{title}{Long-distance continuous-variable quantum key
  distribution with a gaussian modulation}.
\newblock \emph{\bibinfo{journal}{Phys. Rev. A}} \textbf{\bibinfo{volume}{84}},
  \bibinfo{pages}{062317} (\bibinfo{year}{2011}).

\bibitem{SR:prl08}
\bibinfo{author}{Scarani, V.} \& \bibinfo{author}{Renner, R.}
\newblock \bibinfo{title}{Quantum cryptography with finite resources:
  Unconditional security bound for discrete-variable protocols with one-way
  postprocessing}.
\newblock \emph{\bibinfo{journal}{Phys. Rev. Lett.}}
  \textbf{\bibinfo{volume}{100}}, \bibinfo{pages}{200501}
  (\bibinfo{year}{2008}).

\bibitem{LGG:pra10}
\bibinfo{author}{Leverrier, A.}, \bibinfo{author}{Grosshans, F.} \&
  \bibinfo{author}{Grangier, P.}
\newblock \bibinfo{title}{Finite-size analysis of a continuous-variable quantum
  key distribution}.
\newblock \emph{\bibinfo{journal}{Phys. Rev. A}} \textbf{\bibinfo{volume}{81}},
  \bibinfo{pages}{062343} (\bibinfo{year}{2010}).

\bibitem{JKDL:pra12}
\bibinfo{author}{Jouguet, P.}, \bibinfo{author}{Kunz-Jacques, S.},
  \bibinfo{author}{Diamanti, E.} \& \bibinfo{author}{Leverrier, A.}
\newblock \bibinfo{title}{Analysis of imperfections in practical
  continuous-variable quantum key distribution}.
\newblock \emph{\bibinfo{journal}{Phys. Rev. A}} \textbf{\bibinfo{volume}{86}},
  \bibinfo{pages}{032309} (\bibinfo{year}{2012}).

\bibitem{Kikuchi16}
\bibinfo{author}{Kikuchi, K.}
\newblock \bibinfo{title}{Fundamentals of coherent optical fiber
  communications}.
\newblock \emph{\bibinfo{journal}{J. Lightwave Technol.}}
  \textbf{\bibinfo{volume}{34}}, \bibinfo{pages}{157--179}
  (\bibinfo{year}{2016}).

\end{thebibliography}
\end{document}